# Revealing the correlation between real-space structure and chiral magnetic order at the atomic scale

Nadine Hauptmann[1], Melanie Dupé[2], Tzu-Chao Hung[1], Alexander K. Lemmens[1], Daniel Wegner[1], Bertrand Dupé[2], Alexander A. Khajetoorians[1*]

[1] Institute for Molecules and Materials, Radboud University, 6525 AJ Nijmegen, Netherlands

[2] Institut für Physik, Johannes Gutenberg Universität Mainz, D-55099 Mainz, Germany

* Correspondence to: a.khajetoorians@science.ru.nl

**We image simultaneously the geometric, electronic and magnetic structure of a buckled iron bilayer film that exhibits chiral magnetic order. We achieve this by combining spin-polarized scanning tunneling microscopy and magnetic exchange force microscopy (SPEX), to independently characterize the geometric as well as the electronic and magnetic structure of non-flat surfaces. This new SPEX imaging technique reveals the geometric height corrugation of the reconstruction lines resulting from strong strain relaxation in the bilayer, enabling the decomposition of the real-space from the eletronic structure at the atomic level, and the correlation with the resultant spin spiral ground state. By additionally utilizing adatom manipulation, we reveal the chiral magnetic ground state of portions of the unit cell that were not previously imaged with SP-STM alone. Using density functional theory (DFT), we investigate the structural and electronic properties of the reconstructed bilayer and identify the favorable stoichiometry regime in agreement with our experimental result.**

Chiral magnets, exhibiting magnetic skyrmions[1-4], have attracted heavy interest recently, due to their potential application for nano-scale magnetic storage[5, 6]. The dominating mechanism responsible for generating stable non-collinear configurations of the spin is the Dzyaloshinskii-Moriya interaction (DMI), which favors orthogonal configurations of the spin[7, 8]. One key challenge for magnetic data storage is to generate isolated skyrmions with nanometer-scale sizes[9, 10]. Tremendous effort has been made toward this end to stabilize isolated skyrmions at room temperature[5, 11-13], or with length scales down to the nanometer scale at low temperature, utilizing hybrid transition metal interfaces to pin skyrmions at various types of defects[10]. In the latter case, the presence of isolated or arrays of defects raises the question how strong structural relaxations[14], such as surface reconstruction or buckling, simultanously affect the various exchange interactions, geometric structure, and electronic properties, thus affect the magnetic order as well as dynamics of chiral magnetic structures[15, 16].

The most prevalent magnetic imaging technique that can access the magnetization of surfaces and interfaces at the atomic scale is spin-polarizaed scanning tuneling microscopy (SP-STM)[17]. Despite its overwhelming success over the last decade in characterizing surface and interfacial magnetism, SP-STM lacks the capability of solely detecting either the topographic height profile or the magnetic structure, as the tunneling current depends on the structural as well as the electronic and magnetic properties, which cannot easily be decomposed. In contrast, non-contact atomic force microscopy (NC-AFM) gives more accurate structural information as it is sensitive to the forces between the probe and the surface[18]. Combined with magnetic sensitvity, NC-AFM can probe the magnetic exchange forces at the atomic scale (MExFM), yet it has been proven to be difficult to apply[19-22]. Up to now, there has been no demonstration of the combination of these methods to reveal the interplay between real-space and electronic structure, and its effect on magnetic order at the atomic scale.

Here, we use the recently developed combination of SP-STM and MExFM (SPEX)[23] together with *ab-initio* methods to deconvolute the real-space structure of a reconstructed bilayer of Fe on Ir(111) from the electronic and magnetic structure, which exhibits an atomic-scale spin-spiral ground state. We find that, in contrast to previous studies based solely on SP-STM[24], the larger height corrugation of the reconstruction lines as measured by NC-AFM is linked to a lower intensity spin-spiral structure in SP-STM. Furthermore, using a combination of SPEX and manipulation of adsorbed Fe adatoms, we address regions in the unit cell which were previously uncharacterized and extract the complete magnetic structure in relation to the geometric structure of the Fe bilayer. All these measurements provide an essential ingredient for *ab-initio* calculations of such complex structures, and by considering various stochiometries we explain the reconstruction of the Fe bilayer.

Fe layers and single Fe adatoms were deposited *in-situ* on a Ir(111) surface cleaned in UHV and subseqently characterized with an STM/AFM based on the qPlus tuning fork method[25] (PtIr or Fe tips) at $T = 6.3$ K (see section S1 for more details). We performed density functional theory (DFT) calculations using VASP (see section S8 for more details).

Fig. 1 illustrates a characterization of the bilayer Fe on Ir(111) utilizing SPEX imaging, which exhibits a periodic network of reconstruction lines (RLs). This network has been described in a recent work to originate from uniaxial strain relief between the pseudomorphically grown first Fe layer and the second-layer Fe atoms[24]. This RLs network hosts a spin-spiral texture and forms 120° rotational domains, as detailed by magnetic field-dependent measurements in Ref. [24]. The SP-STM image (Fig. 1(a)), acquired with an Fe bulk tip sensitive to the out-of-plane magnetization, shows RLs with a lower (dark) and higher (bright) intensity, marked by a single- and double-headed arrow, respectively. The typical spacing between two bright RLs is 3.3 nm ± 0.2 nm, which is smaller than the recently reported spacing of 5.2 nm[24]. Along the bright and dark RLs, bright protrusions with a periodicity of 1.24 nm ± 0.05 nm are observed, which correspond to the out-of-plane magnetic moments of the cycloidal spin spiral that runs along the RL[24]. We note that the appearance of the spin contrast of the spin spirals strongly depends on the bias voltage (see Fig. S2).

The MExFM image in constant frequency shift mode (Fig. 1b) gives more insight into the height corrugation derived from the RLs. The total force between the Fe bulk tip and the surface responsible for the image contrast consists of various force contributions, which are long- and short-range magnetic forces, van der Waals forces, as well as well short-range chemical forces. We exclude local contributions from electrostatic forces due to the metallic substrate. The arrows in Fig. 1(b) mark the same location of the bright and dark RLs as in Fig. 1(a). Interestingly, the dark RL in the SP-STM image corresponds to a brighter contrast in the MExFM image, which we interpret as a height corrugation of the Fe bilayer. Likewise, the bright RL in the SP-STM image corresponds to a smaller corrugation in the MExFM image. This is counterintuitive and shows that the electronic/magnetic and topographic structure of the bilayer are all strongly convoluted in SP-STM and that it is not possible to unambiguously extract the exact geometric structure from SP-STM alone.

Next, we focus on the vertical relaxations resulting from the RLs. In order to exclude magnetic contributions, we used a non-magnetic PtIr tip attached to the qPlus sensor. Figures 2(a) and (b) show STM and NC-AFM images, respectively, of a close-up view of one RL domain. In agreement with the data acquired with the magnetic Fe tip (Fig. 1), a RL with a larger apparent height in the constant current STM image (Fig. 2(a)) corresponds to a smaller corrugation in the NC-AFM image (Fig. 2(b)) and *vice versa* (indicated by the single- and double-headed arrows). From line profiles (Fig. 2(c)) we extract 4.7 pm ± 1 pm for the larger and 2.2 pm ± 1 pm smaller height corrugation in NC-AFM images, respectively. The given error margins include uncertainties of the piezo constants, the noise extracted from raw data line profiles of NC-AFM images, as well as the variation of the corrugation values for different reconstruction lines. Compared to that, the apparent-height corrugation in the STM image is about one order of magnitude larger than in the NC-AFM image. These findings from combined STM and NC-AFM images indicate that the film exhibits strong vertical relaxations within the unit cell, and thus the Fe film cannot be considered flat[10] necessitating the characterization of SPEX imaging for such strongly relaxed surfaces.

By bringing the PtIr tip closer to the surface, we resolve the previously uncharacterized atomic lattice utilizing constant-current STM imaging (Fig. 2(d)). The atomic lattice spacing in the [1$\bar{1}$0] direction varies between 2.0 Å and 2.7 Å, as indicated by the elongated spots of the FFT (inset). We further find that the atomic lines along the [1$\bar{1}$0] direction do not form rows but rather exhibit a pattern reminiscint of thin transition metal films[26-28], which can be seen by following the atom positions (bright spots) along the horizontal line in Fig. 2(d). Along this line we extract that 16 atoms are located between the two dark RLs. The Fe atoms in the first layer grow pseudomorphically on the Ir(111) atomic lattice with a nearest-atom spacing[29] of 2.72 Å. Considering the spacing of the RLs (3.3 nm ± 0.2 nm) we extract a compression along the [1$\bar{1}$0] direction between 7 and 8% in the Fe bilayer. This is comparable with the compression for the previously suggested model of the atomic arrangement in the Fe bilayer, where a variation of the atom stacking between fcc, hcp and bcc areas has been suggested and a compression of 5 to 6% has been stated[24]. However, the RLs network in our work deviates from recent findings: first, the spacing in between the RLs is smaller in our work. This might be a consequence of different film preparation procedures as described in section S5. Secondly, from NC-AFM and MExFM images we extract a height corrugation up to 5 pm, which has not been described in the previous work in which the structure was considered to be planar. Using SPEX imaging, we can obtain complementary information, which provides direct information on the vertical relaxation for strained film structures providing necessary input for *ab-initio* methods.

In Fig. 3(a), a simulated STM image is presented along with the atomic structure of the reconstructed over-stoichiometric (10:9) surface, which is energetically the most favorable. In total, we calculated reconstruction energies of five different surface structures, of which two form a reconstructed domain pattern and the remaining three are pseudomorphic homogeneous surfaces with bcc-like stacking as well as fcc and hcp stacking of the Fe(top)-layer (see section S8 for more details). The STM image (Fig. 3(a)) was obtained from the electron density calculated 100 pm above the surface layer arising from the electronic states at the Fermi energy, which is proportional to STM current intensities. The Fe(top)-atoms at the surface form a periodic structure (orange atoms), where two differently oriented bcc domains are tilted against each other and twisted with respect to the Fe(bottom)-layer (blue atoms). As a consequence, the atomic rows are not straight but follow a zig-zag pattern as indicated in Fig. 3(a). The domain boundaries are formed by lines of surface atoms Fe(top) possessing fcc or hcp stacking, which appear bright in the STM image, whereas the interior of the bcc domains appears dark.

Fig. 3(b) shows a more detailed analysis of the atomic and electronic structure along the two atomic rows indicated in (a). The relative heights of the Fe(top) atoms are shown in orange and red. The reconstructed surface exhibits a strong buckling with two different types of maxima in both RLs: hcp atoms are higher than fcc atoms by up to 3.4 pm. In blue colors, the electron densities along the same rows obtained at height of 100 pm above the surface (as in the simulated STM image) are given. Counterintuitively, the higher hcp regions at distances 0.0, 2.4 and 4.8 nm possess lower electron densities than the lower fcc regions at distances 1.2 and 3.6 nm. The higher hcp-RL appears slightly darker in the STM image than the lower fcc-RL. The low-lying atoms within the bcc-domains appear darkest. Altogether, our simulated STM image of the over-stoichiometric (10:9) surface agrees well with the findings from atomically resolved STM as well as combined STM/NC-AFM images (Fig. 2). In particular, the existence of hcp and fcc RLs, with different geometrical heights, will have consequences for the magnetic structure as the DMI is strongly dependent on the surface geometry. Our results also show that STM imaging alone cannot reveal the real-space structure, as it convolutes the electronic structure and the surface relaxation. In contrast, NC-AFM can reveal the latter independent of the surface electron density which is essential for consideration of complex magnetic structures[10].

We note that the observed domain width (and bilayer stoichiometry) is determined by the supercell size, which is limited by computational cost. Thus, different periodicity lengths of the domain structure are possible and depend presumably crucially on the preparation conditions of the bilayer, as evidenced by the different RL spacings found in our and the previous work (see section S6). The main features of the reconstructed surface structures, however, will remain unchanged: 1. prevalence of bcc domains and 2. tilt-domain boundaries formed by fcc and hcp atoms resulting in pronounced buckling of the surface layer.

Now that the structural model has been established, we finally analyze the magnetic structure of the Fe bilayer using SPEX in combination with single adatom manipulation to characterize regions of the unit cell previously uncharacterized. A close-up view of one domain of the spin-spiral network is shown in Figs. 4(a) and (b). The SP-STM image agrees with previous measurements[24]. In addition, however, we also observe the spin spiral in the MExFM image, which can be clearly discerned on the RL with the lower corrugation, and is also resolvable on the RL with larger corrugation, as also seen in the overview MExFM image in Fig. 1. While the SPEX images reveal the out-of-plane magnetization on the RLs, we cannot conclude what the magnetization of the atoms between the RLs are, due to the inherent topographic and electronic contrast variation. Therefore, we utilize a deposited Fe adatom, and manipulate its position with respect to the surface magnetic unit cell to infer the magnetization[30] (for details see section S5). This is done in the following manner: (1) additional Fe adatoms are evaporated onto the surface, (2) one adatom is manipulated along a row in between a dark and bright spin spiral, and the changes in spin contrast between the moved atom and a reference atom (remaining at a fixed position) is compared. We presume that the deposited adatoms are strongly exchange-coupled to the surface and electronically equivalent. Therefore, their intensity reflects the change in the magnetic state of the underlying spin spiral without strongly perturbing the underlying magnetic state as long as the coverage is dilute. Figures 4(c) and (d) show SP-STM images of an Fe adatom (white circles) that has been manipulated in between the area of a bright and dark RLs. We reproducibly moved an individual Fe adatom over about 100 different positions (for details see section S5). As seen from Figs. 4(c) and (d), the apparent height of the Fe adatom varies depending on the position.

To systematically investigate the apparent-height variation of the Fe adatom, we consider the apparent-height difference $\Delta h$ between the Fe adatom and the reference Fe atom (R) whose apparent height does not change during the entire manipulation process. Figure 4(e) shows the dependence of $\Delta h$ along the *y* direction (as defined in Fig. 4(c)). The apparent height oscillates at different positions along the spin spiral with the same periodicity as the spin spiral (between 1.2 and 1.3 nm). As it can be seen from Figs. 4(c) and (d), the apparent-height variation of the Fe adatom is phase-shifted by 180° with respect to the contrast of the spin spiral. Along the direction orthogonal to the *y* direction, $\Delta h$ does not vary significantly (see section S5 for details). We conclude from our data that the out-of-plane magnetic moment of the adatom alternates along the $[11\bar{2}]$ direction while the variation along the $[1\bar{1}0]$ is small. Our observation is in agreement with a previously suggested magnetic structure[24], if we assume that the Fe adatom exhibits an inversion of the spin polarization with respect to the underlying Fe bilayer[31, 32].

In summary, we used the new SPEX imaging technique together with *ab-initio* DFT calculation to experimentally delineate the interplay between surface reconstruction and magnetic structure of the spin-spiral network in Fe bilayers on Ir(111) for the first time. We show that the reconstruction lines are subject to a substantial vertical corrugation ($\Delta z \approx 4$ pm), resulting from the energetically favorable over-stoichiometric surface, which we measured by NC-AFM and confirmed using DFT. Our results illustrate that strong relaxation effects due to defects or strain, cannot be extracted from SP-STM measurements alone and may be an important consideration for *ab initio* calculations of the magnetic structure as well as in thin magnetic films[10]. These strong structural relaxations may have an important effect on the exchange interactions and potentially on the DMI that stabilize the magnetic texture of the bilayer by modifying the interfacial hybridization, and therefore require careful consideration.

**Acknowledgements**

We would like to thank Kirsten von Bergmann, Andre Kubetzka, Jairo Sinova and Stefan Heinze for useful discussions. We would like to acknowledge financial support from the Emmy Noether Program (KH324/1-1) via the Deutsche Forschungsgemeinschaft, and the Foundation of Fundamental Research on Matter (FOM), which is part of the Netherlands Organization for Scientific Research (NWO), and the VIDI project: 'Manipulating the interplay between superconductivity and chiral magnetism at the single atom level' with project number 680-47-534 which is financed by NWO. NH and AAK also acknowledge support from the Alexander von Humboldt Foundation via the Feodor Lynen Research Fellowship. MD and BD gratefully acknowledge the computing time granted on the supercomputer Mogon at Johannes Gutenberg University Mainz (hpc.uni-mainz.de). BD and MD also acknowledge financial support from the Alexander von Humboldt Foundation and the Transregional Collaborative Research Center SFB/TRR 173 "Spin+X" from the Deutsche Forschungsgemeinschaft.

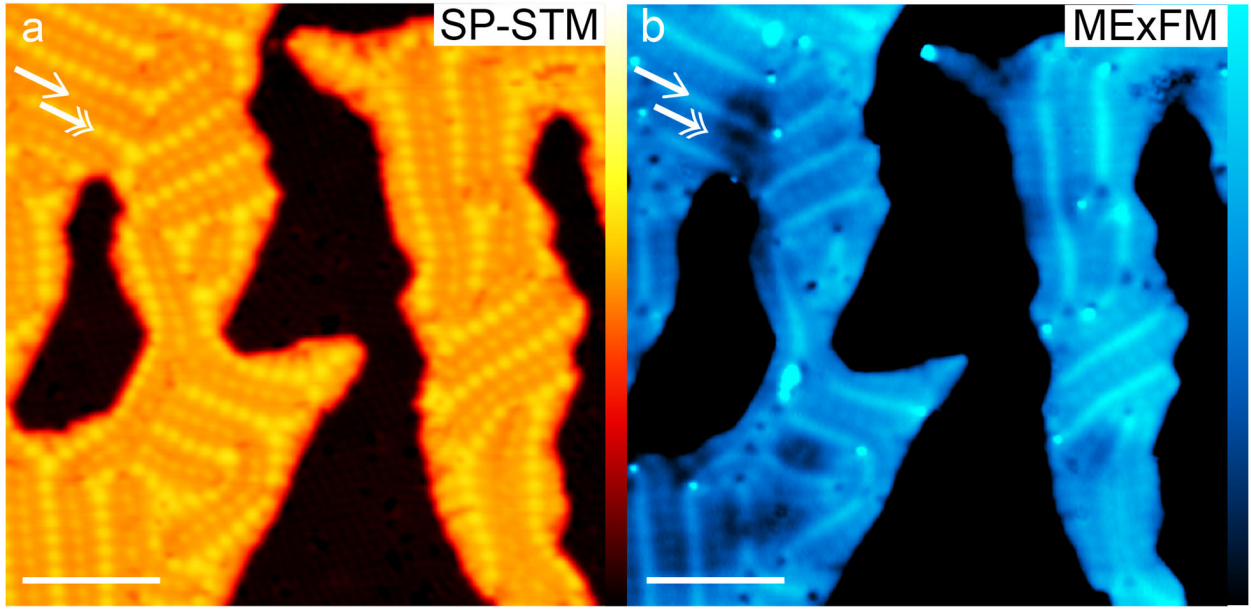

Figure 1: Large-scale (a) constant current SP-STM image ($V_s$ = 50 mV, $I_T$ = 100 pA, Fe tip) and constant frequency shift MExFM image ($\Delta f_{set}$ = −12 Hz, $z_{mod}$ = 100 pm, $V_s$ = 0.1 mV, Fe tip). Both images: 43.5 x 43.5 nm². Scale bars are 10 nm. Color scales are (a) 0.4 to 500 pm and (b) 220 to 240 pm.

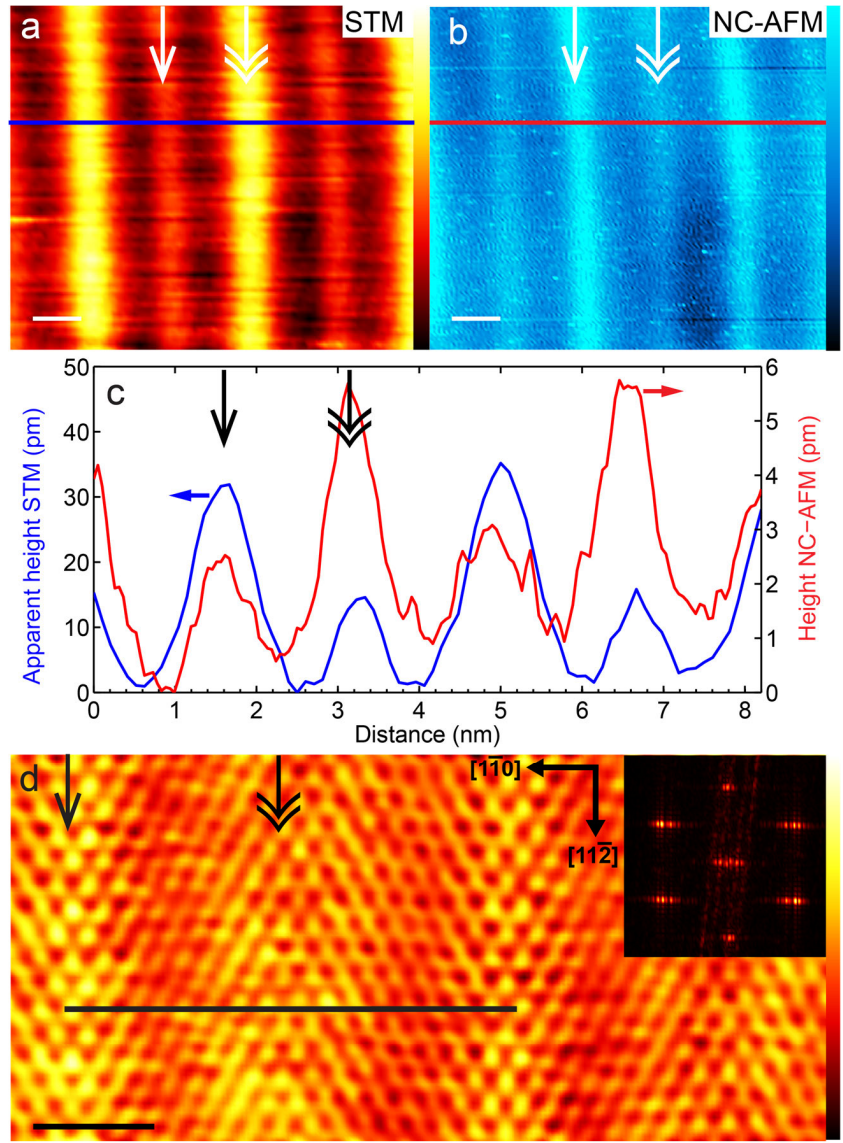

Figure 2: (a) Constant-current STM ($V_s$ = 50 mV, $I_T$ = 100 pA, PtIr tip) and constant frequency shift NC-AFM images (flattened by an offset, $\Delta f_{set} = -40$ Hz, $z_{mod}$ = 100 pm, $V_s$ = 0.1 mV). (c) Line profiles of the STM and NC-AFM images along the horizontal lines as indicated in (a) and (b). The NC-AFM has been smoothed prior to taking the line profile by a Gaussian filter with 2 points. (d) Atomic resolution STM image (flattened by an offset, Gaussian smoothed by 2 points, $V_s$ = 1 mV, $I_T$ = 48 nA, PtIr tip), inset shows the FFT of (d). Sizes: (a) 8.5 x 7.3 nm², (b) 8.5 x 7.3 nm², (d) 6.8 x 3.3 nm². Scale bars are 1 nm for all images. Color scales are (a) 0 to 54 pm and (b) 1 to 11 pm, (d) 0 to 28 pm. The arrows indicate equivalent positions in all images.

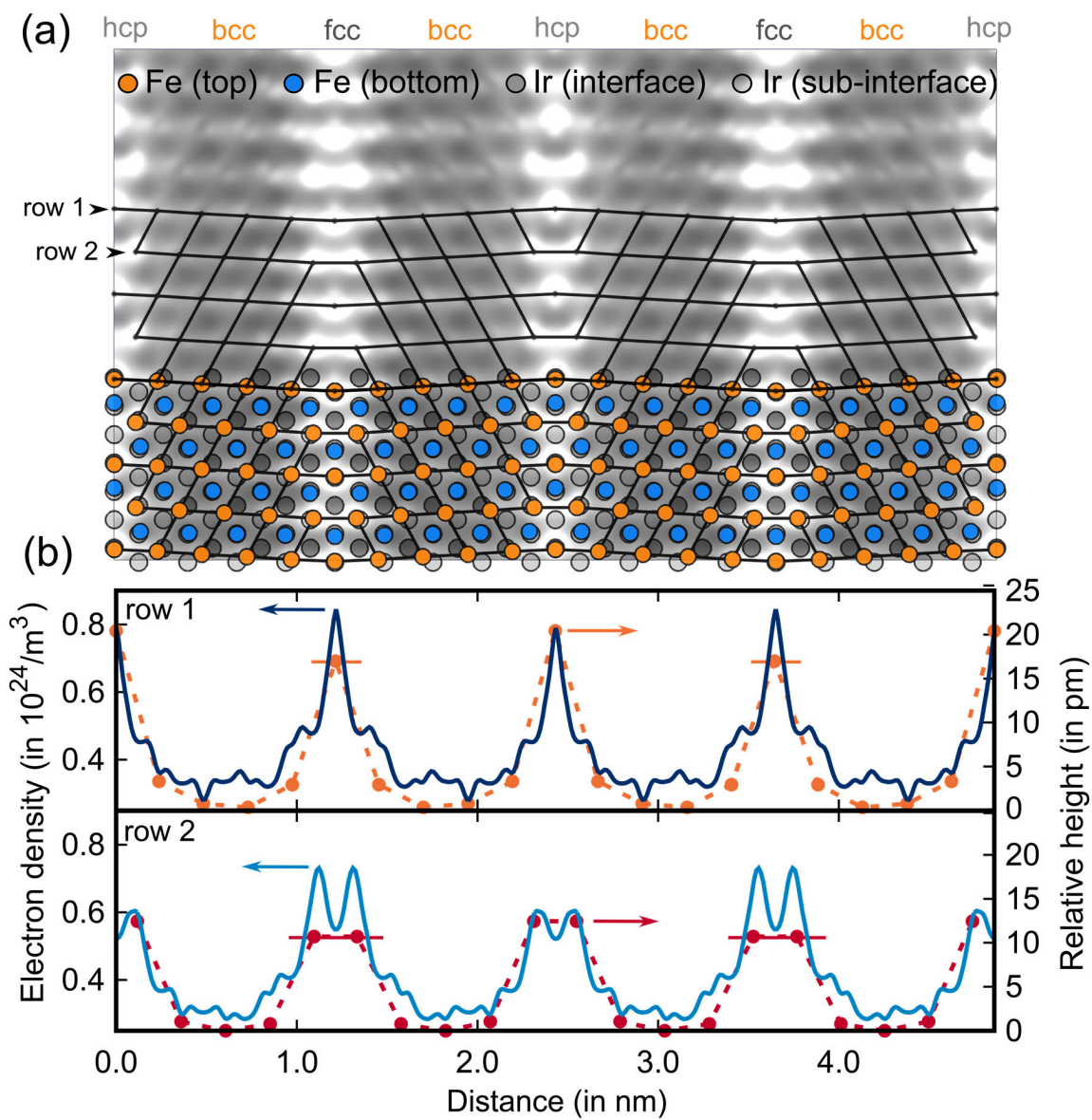

Figure 3: Characterization of the reconstructed over-stoichiometric (10:9) surface from *ab-initio* supercell calulations. In (a), a simulated STM image is presented along with the atomic structure of the reconstructed surface. In (b), a more detailed analysis of the atomic and electronic structure along the two atomic rows indicated in (a) is given.

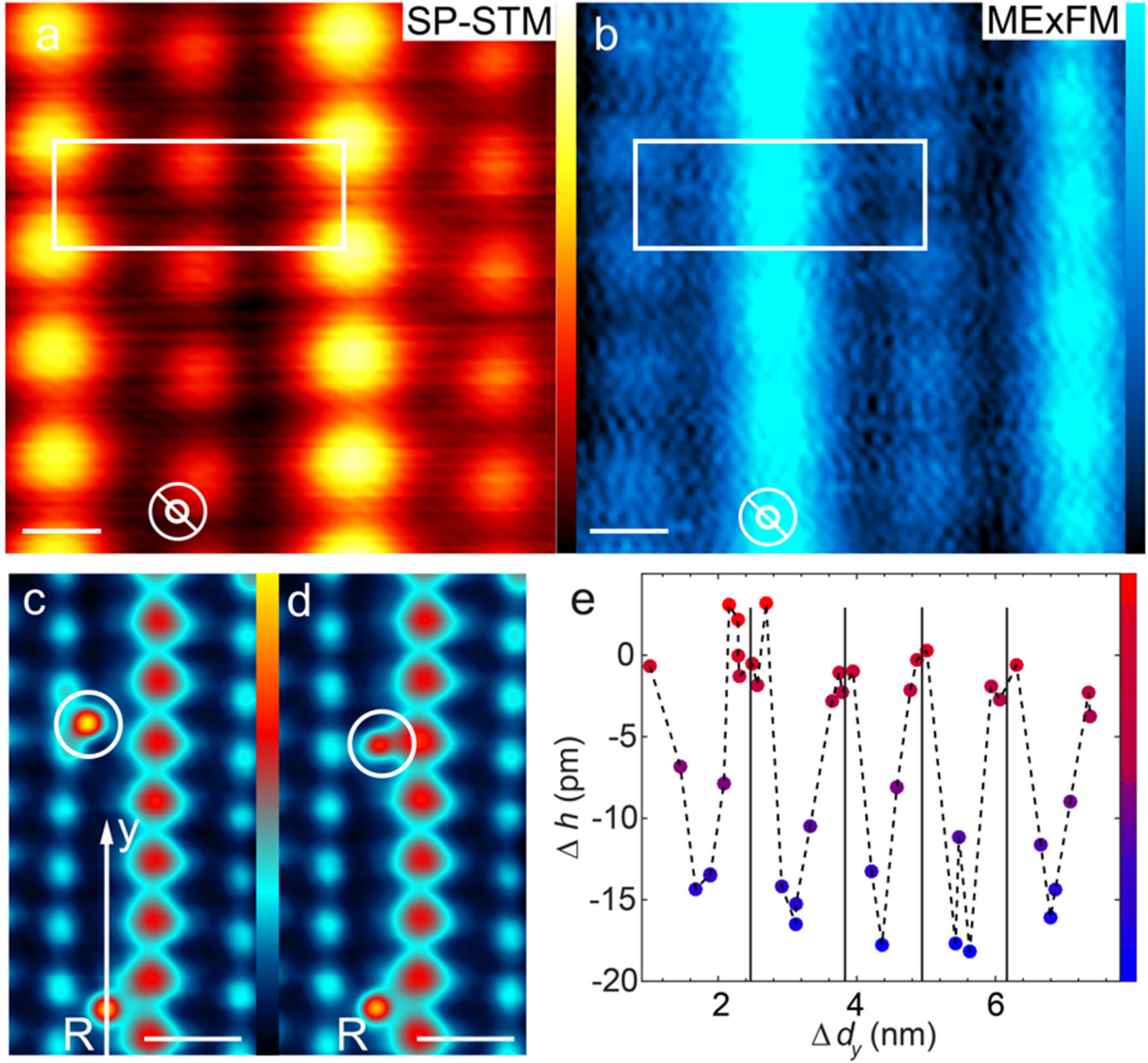

Figure 4: (a) Constant-current SP-STM ($V_s$ = 50 mV, $I_T$ = 100 pA, Fe tip) and constant frequency shift MExFM image (flattened by an offset, smoothed by a Gaussian filter with 2 points, $\Delta f_{set}$ = −19 Hz, $z_{mod}$ = 100 pm, $V_s$ = 0.0 mV, Fe tip). Scale bars are 1 nm. The tip magnetization is out-of-plane as indicated by the symbol. (c) and (d) Constant-current SP-STM images (Fe tip, $V_s$ = 50 mV, $I_T$ = 100 pA, smoothed by Gaussian filter 2 points), scale bars are 2 nm. Color bar: (a) 0 to 56 pm, (b) 0 to 5 pm, (c) and (d): 0 to 76 pm. The Fe adatom (white circle) was moved in between the dark and bright spin spirals. The reference atom R rests at the same position. (For a video of the entire data set, see section S5). (e) Apparent height $\Delta h$ of the adatom versus the distance $\Delta d_y$ to the reference atom R when moved along a line in between the dark and bright spin spiral. The y direction is defined in (c).

# Revealing the correlation between real-space structure and chiral magnetic order at the atomic scale

Nadine Hauptmann[1], Melanie Dupé[2], Tzu-Chao Hung[1], Alexander K. Lemmens[1], Daniel Wegner[1], Bertrand Dupé[2], Alexander A. Khajetoorians[1*]

[1] Institute for Molecules and Materials, Radboud University, 6525 AJ Nijmegen, Netherlands

[2] Institut für Physik, Johannes Gutenberg Universität Mainz, D-55099 Mainz, Germany

[*] Correspondence to: a.khajetoorians@science.ru.nl

**Section S1: Experimental details**

The experiments were carried out in a commercial ultra-high vacuum low-temperature STM/AFM system from CreaTec with a base pressure of about $2\cdot10^{-10}$ mbar. The Ir(111) surface was prepared by repeated cycles of Ne$^+$ sputtering and annealing ($T \sim 1800$ K) in an oxygen atmosphere ($p \sim 4\cdot10^{-6}$ mbar) followed by a final flash to 1800 K. The Fe layers (about 1.2 ML) were deposited from an e-beam evaporator onto the Ir(111) surface, which was kept at room temperature and subsequently annealed ($T \sim 630$ K). This results in the formation of multilayer Fe islands (see section S2), in which the first layer exhibits both hcp and fcc stacking[1,2]. Afterward, the sample was transferred *in situ* into the cryogenic STM/AFM, which operates at a base temperature of $T = 6.3$ K. Single Fe adatoms were deposited with the sample remaining inside the microscope and being held at 7 K. Combined SP-STM and MExFM (SPEX) measurements were performed using CreaTec's Besocke-type STM/AFM head. For AFM measurements a non-contact frequency-modulation mode was used utilizing a tuning fork-based qPlus sensor[3] with its free prong oscillating at its resonance frequency $f_0 \approx 27.7$ kHz. The force is indirectly measured by the shift of the resonance frequency $\Delta f$ with oscillation amplitudes $z_{\text{mod}}$ (half the peak-to-peak value) between 70 pm and 110 pm. For (SP-)STM measurements a Fe or a PtIr bulk tip was glued to the free prong of the tuning fork. All magnetic images in this work are acquired with an out-of-plane magnetized tip. Details on the characterization procedure of the magnetic tip can be found in the Supporting Information of Ref.[4].

**Section S2: Large-scale SP-STM image**

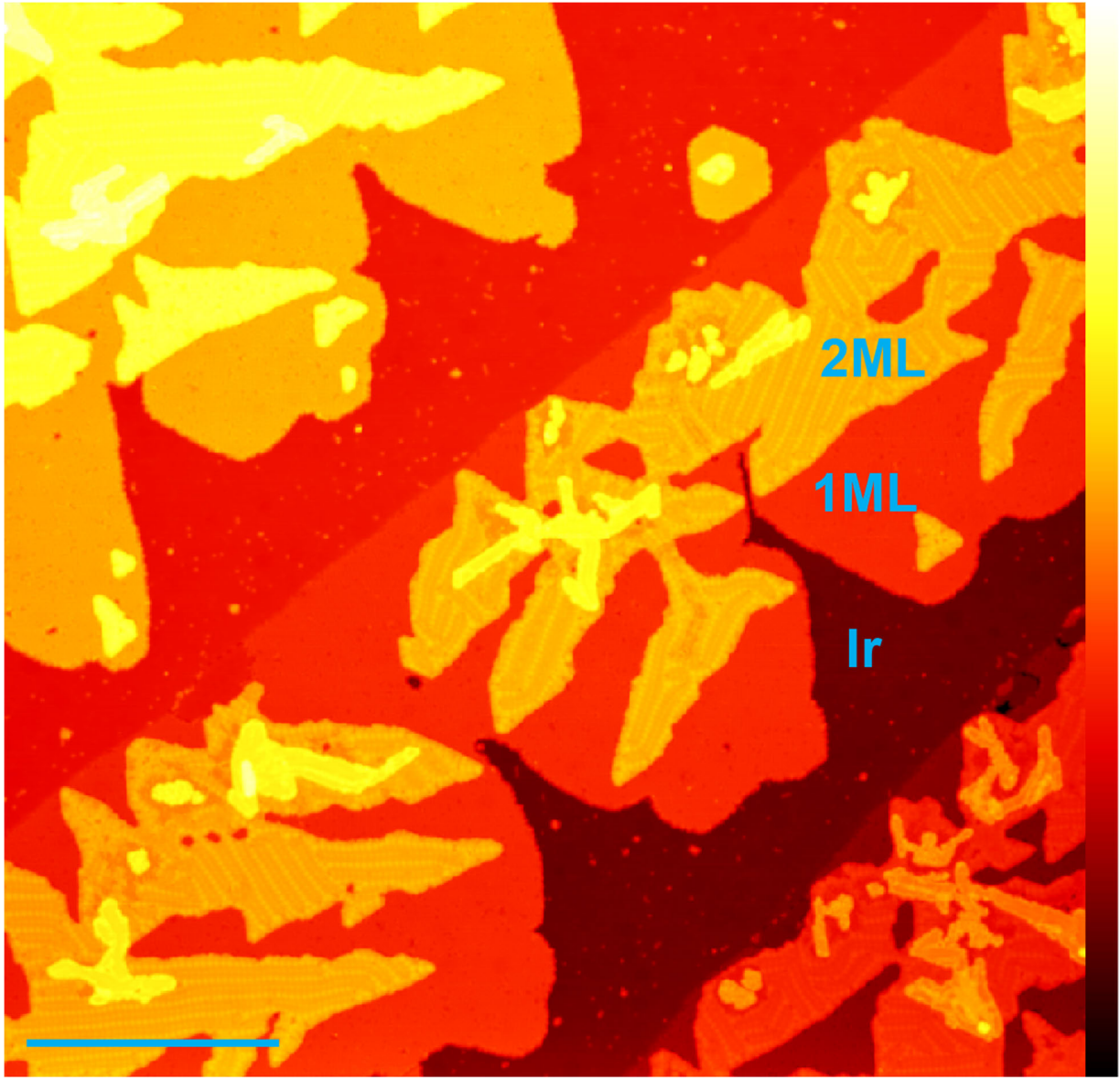

**Figure S1:** Large-scale constant-current SP-STM image of Fe multilayer islands on Ir(111) ($V_s$ = 50 mV, $I_T$ = 100 pA), 215 x 215 nm$^2$. Color scale: 0 to 1.4 nm. Scale bar is 50 nm.

**Section S3: Voltage-dependent SP-STM imaging**

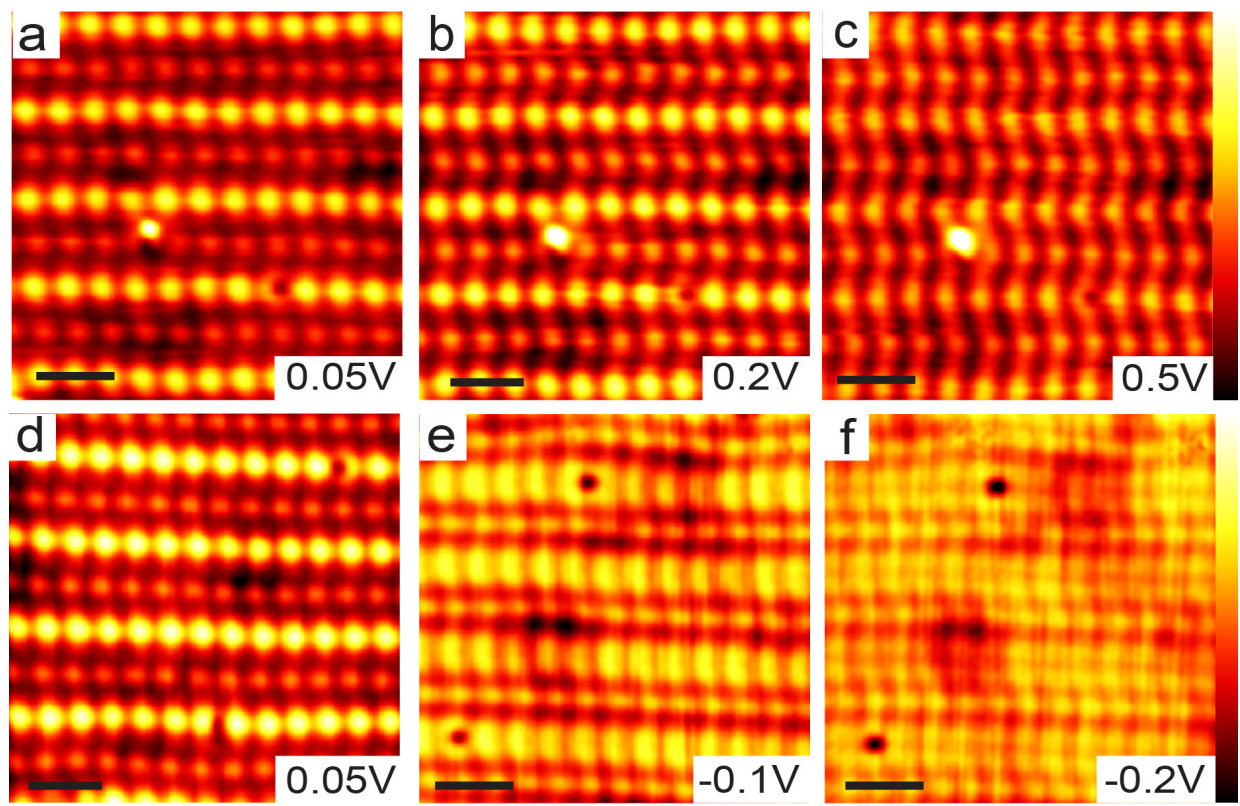

**Figure S2:** Voltage-dependent constant-current SP-STM images ($I_T = 100$ pA, smoothed with Gaussian filter of 2 points): (a) to (c): 15 x 15 nm$^2$, (d) to (f): 16 x 16 nm$^2$. Scale bars are 3 nm in all images. Color scales: (a) 0 to 56 pm, (b) 0 to 70 pm, (c) 0 to 58 pm (d) 0 to 57 pm, (e) 0 to 36 pm, (f) 3 to 35 pm

## Section S4: Constant-height imaging

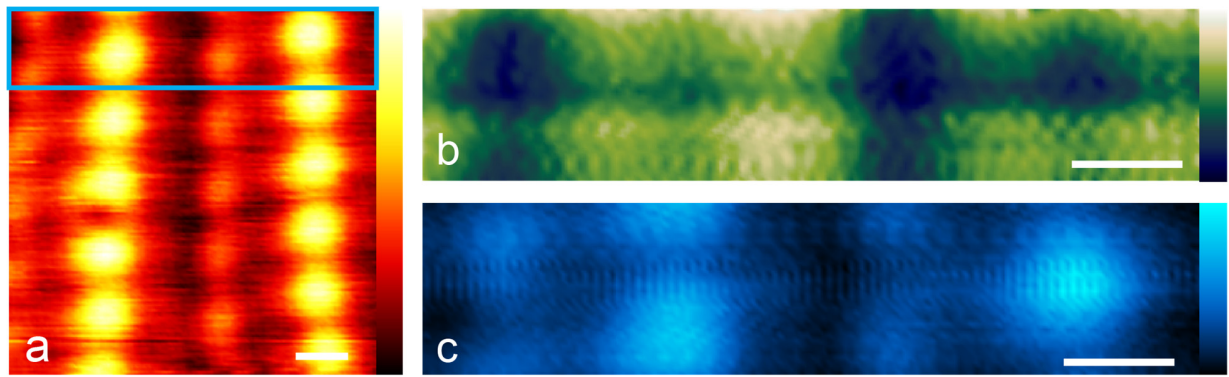

**Figure S3:** Constant-current SP-STM topography ($V_s$ = 50 mV, $I_T$ = 100 pA) (b) Frequency shift and (c) current measured simultaneously ($z_{mod}$ = 70 pm, $V_s$ = 0.1 mV, I/V converter gain: $1 \cdot 10^7$) at a constant tip-sample distance at the position indicated by the blue box in (a). The current-feedback loop was opened at $V_s$ = 50 mV, $I_T$ = 100 pA and the tip was brought closer by 0.42 nm. Sizes: (a) 7 x 7 nm², (b) and (c): 7 x 1.6 nm². Scale bars are 1 nm in all images. (b) and (c) were smoothed by a 2-point Gaussian filter and a global plane was subtracted. Color scales: (a) 0 to 62 pm, (b) -1.20 to 1.2 Hz (c) -0.5 to 1.3 nA.

## Section S5: Fe adatom manipulation

The movie *spinspiral.mp4* shows the subsequently acquired images after movement of the Fe adatom in between a dark and bright spin spiral.

The *movie 2D_map_apparent_height.mp4* shows the movement of two Fe adatoms from which the apparent height map in Fig. S4 was extracted.

The manipulation process was done as follows: First, an SP-STM image at $U$ = 50 mV and $I_T$ = 100 pA with an out-of-plane tip is acquired. Then, the tip is moved above the adatom that should be manipulated. The bias voltage is decreased to 1.5…2mV. After that, the current is increased in constant-current mode to the range from 15 to 30 nA. The magnitude of the current is dependent on the sharpness of the tip. Then, the tip is moved with a speed of about 1 nm/s toward the desired position and the apparent-height profile is monitored, typically showing indications of the movement of the adatom[5]. After the tip has been moved to the desired position, the current and bias voltage have to be switched back to the previous parameters. Finally, a SP-STM image is acquired to confirm that the adatom has moved to the dedicated position as well as to check that neither the geometric nor the magnetic structure of the tip has been altered. The map of the different adatom positions in Fig. S4 has been performed without any change of the tip.

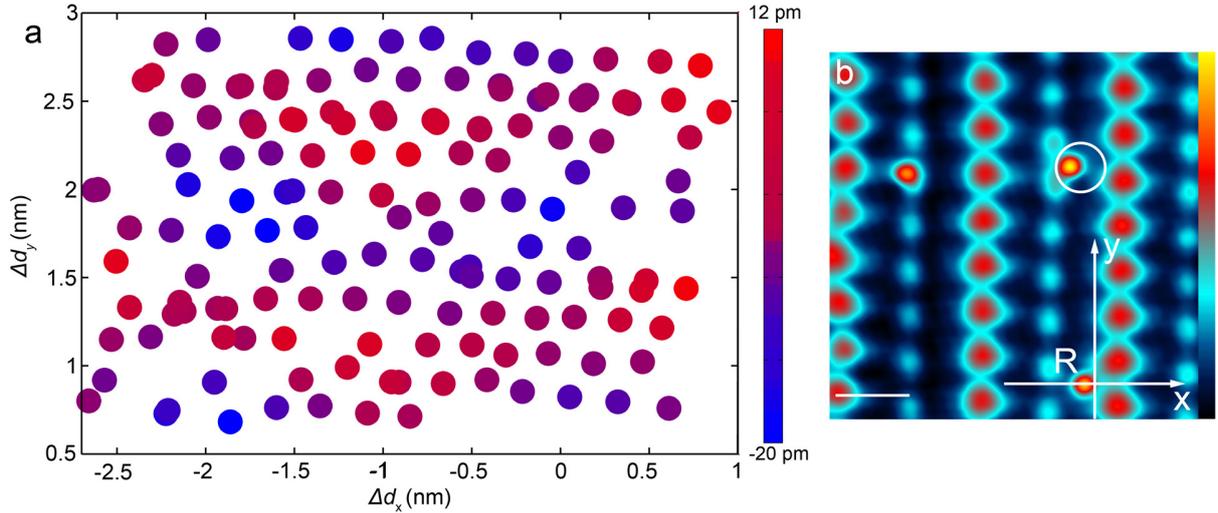

**Figure S4:** (a) 2D apparent-height ($\Delta h$) map along the x and y direction for various positions of the Fe adatom on the Fe bilayer. $\Delta h$ is measured with respect to the apparent height of the reference adatom R. The coordinate system is defined in (b). (b) Constant current SP-STM ($V_s$ = 50 mV, $I_T$ = 100pA, Fe tip). Scale bars is 1 nm. Color scale: 0 to 76 pm.

**Section S6: Analysis of the peridocity of the spin spirals**

We analyzed the different periodicities of the reconstruction lines (RLs) using SP-STM, combined STM/NC-AFM and SPEX images. First, we determined the spacings from 5 SP-STM images (one shown in Fig. S5(a)), acquired with different tips. We obtain same spacings from MExFM data. The typical spacing between two bright (as well as dark) RLs is 3.3nm ± 0.2 nm. The terms dark and bright refer to the appearance of the RLs in the (SP-)STM images showing a lower and higher intensity, respectively. Our extracted distances are smaller than the recently reported spacing of 5.2 nm in Ref. [6]. A possible explanation for the deviation might be the different preparation procedure: while the Fe was grown onto the Ir(111) surface at elevated temperatures in the previous work, here it was deposited on the surface at room temperature and subsequently annealed to about 350°C to form Fe islands. We note, however, that for two islands we observed areas with a spacing of 5 nm.

The height corrugation as extracted from two NC-AFM images acquired with different tips of the dark and bright RLs is 4.7 pm ± 0.3 pm and 2.2 pm ± 0.5 pm, respectively. For the areas with a larger RL spacing (5 nm), we find that the height corrugation as extracted from a MExFM image for the dark appearing RL is smaller: it is 1.8 pm ± 0.3 pm compared to 4.5 pm for the corrugation of the RLs with the typical spacing. This indicates that the vertical strain is relieved more by a larger RL spacing and thus results in a smaller vertical corrugation. The periodicity of the out-of-plane magnetic position along the spin spiral is 1.24 nm ± 0.05 nm, which is the same for the bright and dark RLs.

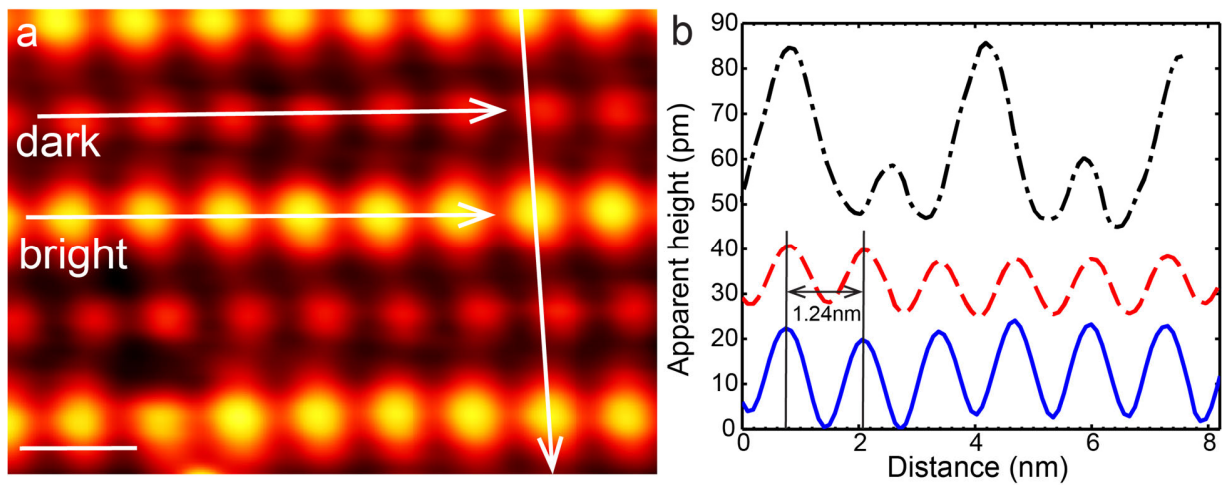

**Figure S5:** (a) Constant-current SP-STM topography ($V_s$ = 50 mV, $I_T$ = 100 pA, Gaussian smoothed by 2 points). Size: 11 x 8nm². Scale bar is 2 nm. Color scale: 0 to 67 pm. (b) Line profiles along the directions indicated in (a): solid blue: along the bright RL; dashed red: along the dark RL; dashed-dotted: orthogonal to the RLs. The FWHM of the out-of-plane components of the magnetization of the spin spirals (bright protrusions) on the RLs is different for spin spirals on dark and bright RLs. For the bright RL the FWHM is 0.93 nm ± 0.05 nm, for the dark RL it is 0.62 nm ± 0.05 nm. The typical spacing between two dark (as well as bright) RLs is 3.3 nm ± 0.2 nm.

## Section S7: Voltage dependence of SPEX images

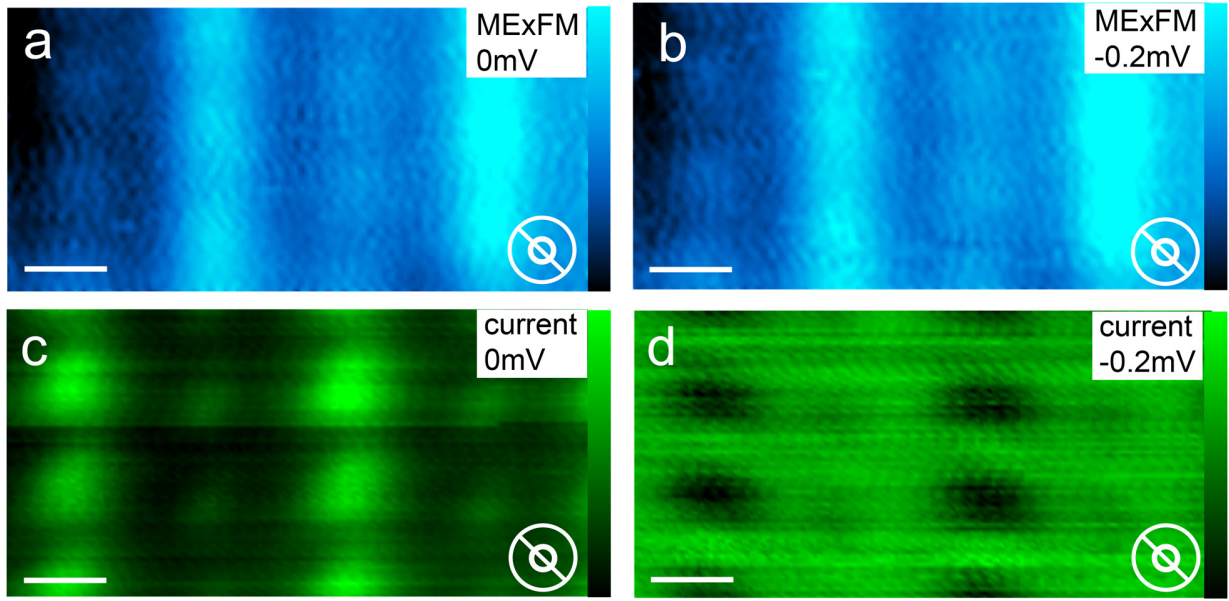

**Figure S6:** (a) and (b): constant frequency shift MExFM image, (flattened by an offset, smoothed by a Gaussian filter with 2 points, $\Delta f_{set} = -19$ Hz, $z_{mod} = 100$ pm, (a): $V_s = 0.0$ mV, (b): $V_s = -0.2$ mV, Fe tip). (c) and (d) simultaneously measured current (I/V converter gain: $1\cdot 10^7$) with (a) and (b), respectively. Sizes of all images: 7 x 3.5 nm$^2$, scale bars are 1 nm. Color scales: (a) 1.1 to 9.1 pm, (b) 1.1 to 9.1 pm, (c) 200 to 570 pA, (d) -21 to -167 pA.

## Section S8: Details of the *ab-initio* calculations

**Computational Method**

We performed density functional theory (DFT) calculations using the Vienna *ab-initio* simulation package VASP[7-10]. We employed projector augmented plane waves (PAW)[11, 12] with the GGA exchange correlation functional in the parameterization of PBE[13]. As cut-off energy of the plane waves 300 eV were chosen. A Γ-centred Monkhorst-Pack mesh[14] of 1x5x1 k-points was used containing three k-points in the irreducible part of the Brillouin zone for the 9x1x1 supercell in the orthorhombic setting, which contains 18 Ir atoms per substrate layer. The in-plane lattice constant was kept constant at 2.70 Å. The two Fe-layers were allowed to relax until residual forces were smaller than 0.01 eV/ Å, while the positions of the Ir atoms were kept fixed. Magnetic moments were initialized in the ferromagnetic order. The crystal structures were visualized using the software VESTA[15].

We considered several 9x1x1 supercells with different bilayer stoichiometries containing about 160 atoms in the symmetric slab geometry of 2Fe|5Ir|2Fe. By varying the Fe content of the top layer (resulting in stoichiometries of 8:9, 9:9 and 10:9), while the Fe bottom layer is kept in the fcc stacking of the Ir(111) substrate, we can mimic different homogeneous and domain surface structures. The starting configurations of the surface structures are given in Fig. S7.

(a) under-stoichiometric (8:9)

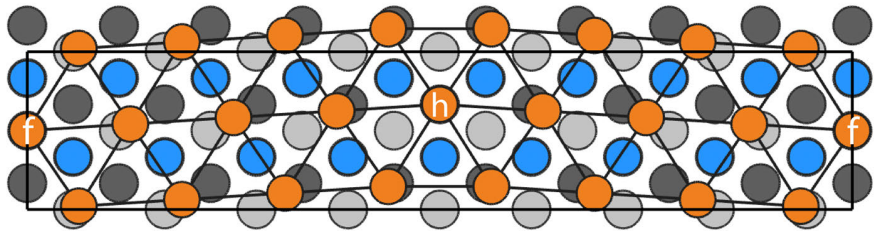

(b) stoichiometric (9:9)

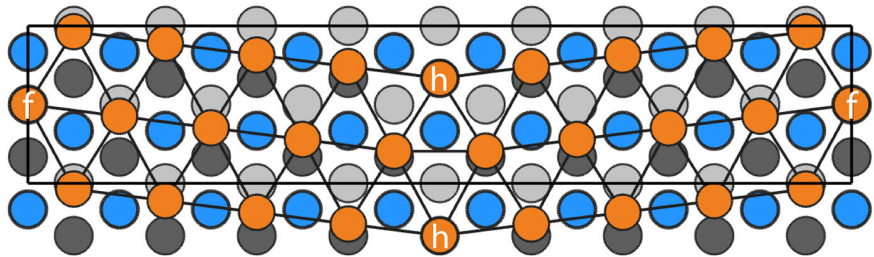

(c) over-stoichiometric (10:9)

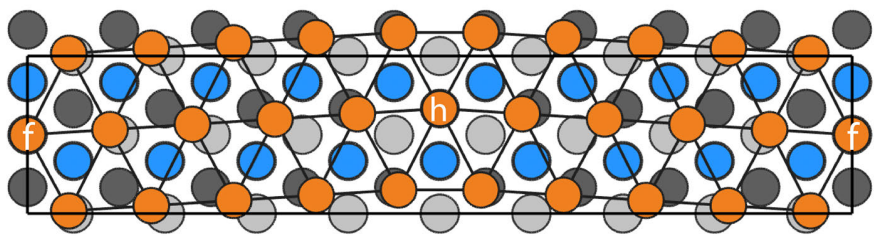

○ Fe (top)   ● Fe (bottom)   ● Ir (interface)   ○ Ir (sub-interface)

**Figure S7:** Supercells of structures before relaxation: Top views of the (a) under-stoichiometric (8:9), (b) stoichiometric (9-9) and (c) over-stoichiometric (10:9) surface structures. The simulated supercell is indicated in black. Fe(top) atoms in fcc (f) and hcp stacking (h) are marked.

In order to gain a deeper understanding of the stability of the over-stoichiometric surface structure, we calculated reconstruction energies of five different surface structures, of which two form a reconstructed domain pattern and the remaining three are pseudomorphic homogeneous surfaces with bcc-like stacking as well as fcc and hcp stacking of the Fe(top)-layer (Fig. S8). The reconstruction energy quantifies the energy gained upon structural changes of the surface, which can be accompanied by insertion or extraction of Fe atoms, as compared to the most stable stoichiometric surface structure of the bilayer, which was chosen as surface reference.

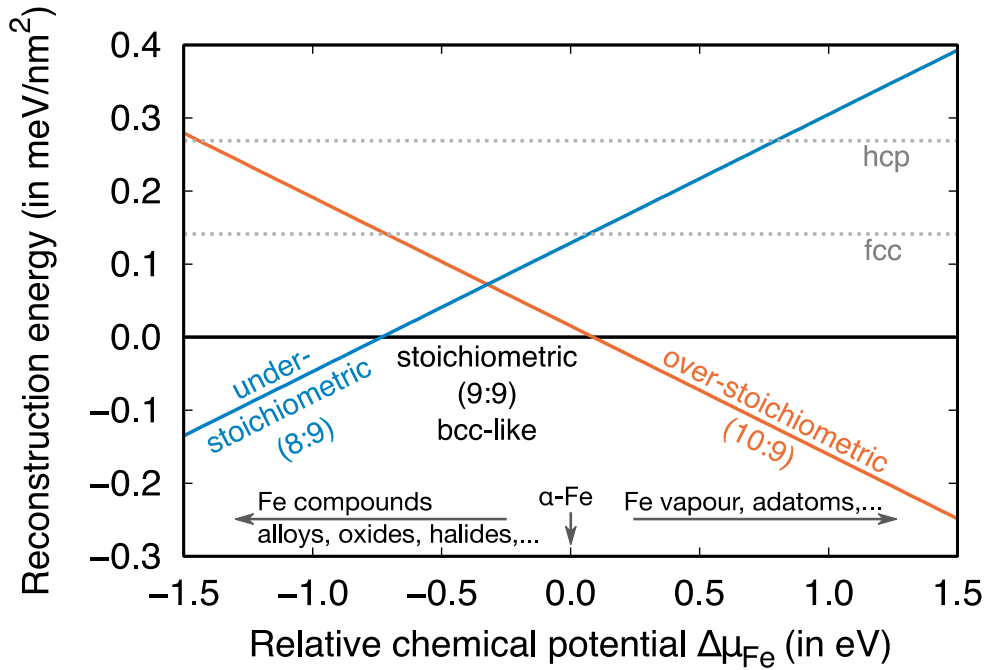

**Figure S8:** Reconstruction energies for different surface structures are given as determined from *ab-initio* supercell calculations. For non-stoichiometric surfaces, the energy gained or lost upon surface reconstruction depends on the energy of the incorporated or extracted Fe atoms. The cohesive energy of α-Fe (ferromagnetic, bcc) is taken as reference state for these Fe atoms ($\Delta\mu_{Fe} = 0$ eV).

Reconstruction energies are calculated from total energies of relaxed structures as follows:

$$\Delta_f E = \frac{1}{A} \left( E_{\text{supercell}} - E_{\text{bcc-like}} - n\Delta\mu_{Fe} \right)$$

where $A$ is the surface area, $E_{\text{supercell}}$ and $E_{\text{bcc-like}}$ are the total energies of the reconstructed supercell and the bcc-like surface structure, and $n$ is the difference of the number of Fe atoms in the surface layer as compared to the stoichiometric surface, it leads to the different slopes of the reconstruction energies.

Depending on the chemical potential of the extra (or extracted) Fe atoms, the reconstruction energy varies, resulting in the stabilization of under-stoichiometric or over-stoichiometric bilayers at negative or positive chemical potential, respectively. The cohesive energy of α-Fe (ferromagnetic, bcc) is taken as reference state for these Fe atoms ($\Delta\mu_{Fe} = 0$ eV). We find for the full range of Fe chemical potentials, that the formation of bcc domains is favored. If the composition is stoichiometric (for $-0.7 < \Delta\mu_{Fe} < 0.1$ eV), the domain structure vanishes and a homogeneous bcc-like layer results. In the chemical potential range typical for the growth conditions of the bilayer, positive values of the relative chemical potential are adopted, which correspond to Fe atoms in the gas phase or adsorbed on the surface, resulting in the formation of the over-stoichiometric surface under equilibrium conditions.

**Analysis of the relaxed structures**

The STM images and charge-density line profiles were obtained from the electron density arising from states at the Fermi energy ($E_{Fermi} \pm 0.1$ eV) at a height of 1 Å (100 pm) above the surface. The simulated STM images of the under-stoichiometric and stoichiometric bcc-like surface structures are presented in Fig. S9 along with the atomic structure after relaxation. The stoichiometric surface relaxes into a homogeneous bcc-like stacking, while the under-stoichiometric surface adopts a bcc-domain structure with Fe(top) atoms in fcc and hcp stacking along the tilt-domain boundaries, similar to the over-stoichiometric surface discussed in the main text.

In Fig. S10, we compare the atomic height profiles of the three stable structures: under-stoichiometric, stoichiometric bcc-like and over-stoichiometric. While the over-stoichiometric surface exhibits a buckling, in the under-stoichiometric surface, trenches can be found forming the reconstruction lines.

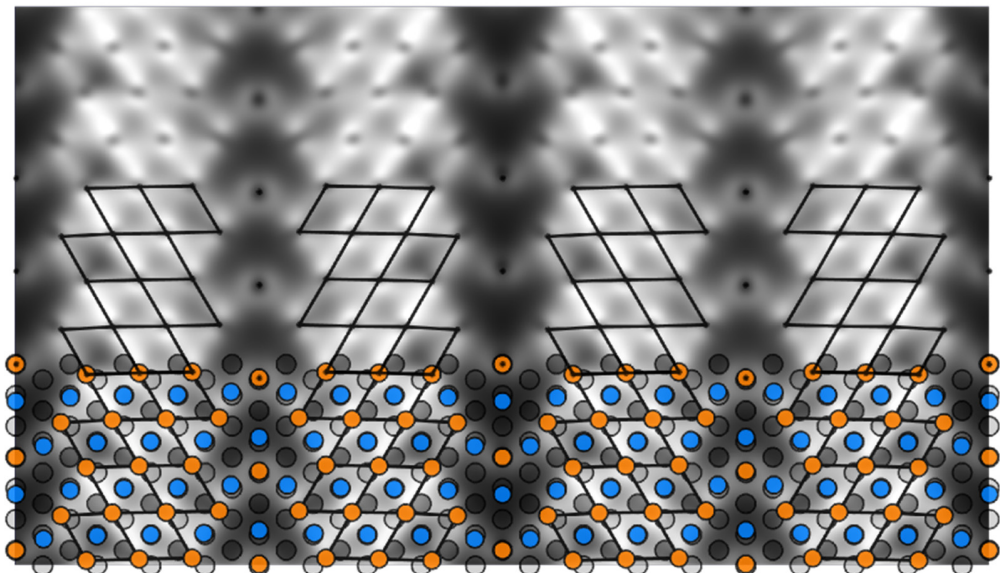

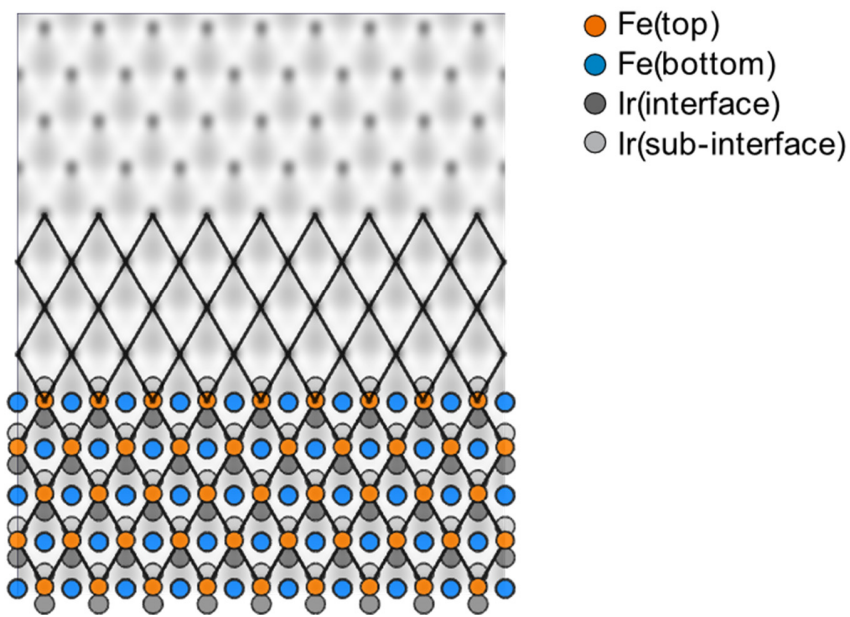

**Figure S9:** Simulated STM images and atomic structures of relaxed supercells: (top) under-stoichiometric (8:9) and (bottom) stoichiometric bcc-like (9:9) surface structures. Color coding of the atoms is the same as in Fig. S7.

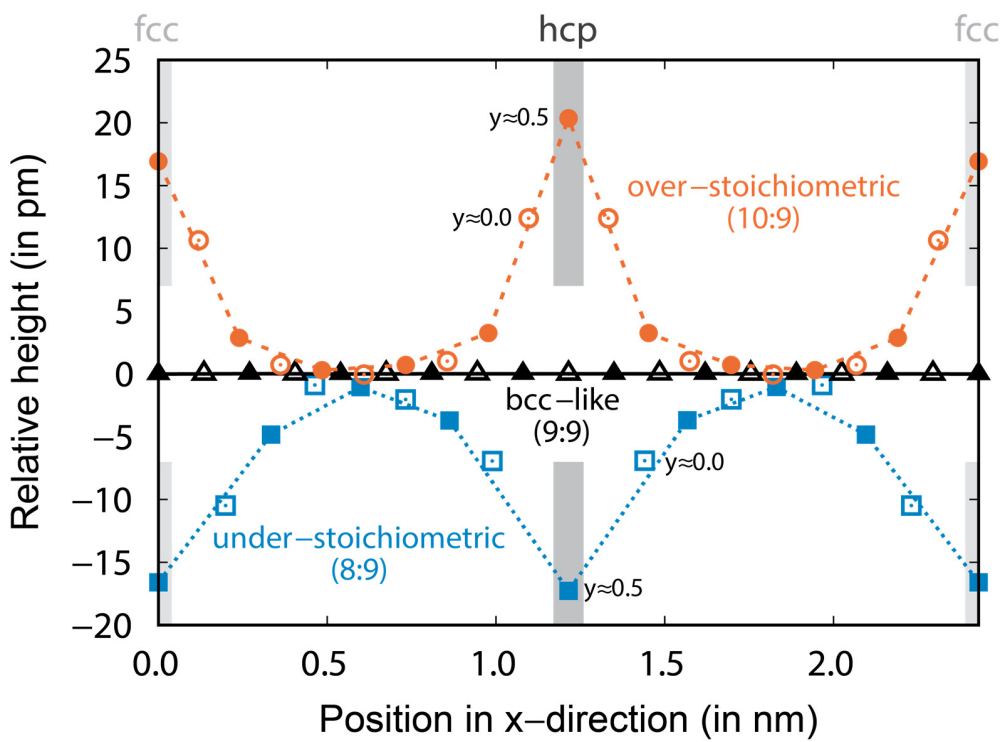

**Figure S10:** Height profiles of stable surface structures: The height profiles were obtained from atomic positions after relaxation for the three stable surface structures: under-stoichiometric (8:9), stoichiometric bcc-like (9:9) and over-stoichiometric (10:9). For the $y$ positions of atoms, please consult Fig. S9. Indicated are the atoms in fcc and hcp stacking in the surfaces with domain structure. Lines are guides to the eye.